\documentclass[twocolumn]{aastex631}

\usepackage{amsmath}
\usepackage{booktabs}
\usepackage{svg}
\usepackage{gensymb}
\usepackage{txfonts}

\begin{document}

\title{Seeing at Timau National Observatory Based on ERA5 Dataset}

\author[0000-0003-1203-2651]{R. Priyatikanto}
\affiliation{Research Centre for Space, National Research and Innovation Agency, Bandung, Indonesia}
\correspondingauthor{R. Priyatikanto (rhor001@brin.go.id)}
\author[0009-0004-0979-7295]{G. P. Putri}
\author[0009-0007-7908-7952]{C. Y. Yatini}
\affiliation{Research Centre for Space, National Research and Innovation Agency, Bandung, Indonesia}
\author{I. Rusyda}
\affiliation{Research School of Astronomy \& Astrophysics, Australian National University, Canberra, Australia}
\author{E. I. Akbar}
\affiliation{Astronomy Research Group, Faculty of Mathematics and Natural Sciences, Institut Teknologi Bandung, Bandung, Indonesia}
\author[0000-0003-1003-1583]{A. G. Admiranto}
\author{F. Noor}
\author{S. Maryam}
\author{Elyyani}
\affiliation{Research Centre for Space, National Research and Innovation Agency, Bandung, Indonesia}
\author{A. Rachman}
\affiliation{Timau National Observatory, National Research and Innovation Agency, Bandung, Indonesia}


\begin{abstract}
Understanding the seeing conditions is crucial for astronomical observations using a ground-based telescope. This study analyzes long-term atmospheric data (2002–2021) from the ERA5 dataset to assess the seeing conditions at the new Timau National Observatory in Indonesia, which will house a 3.8-meter optical telescope. While the ERA5 dataset shows good agreement with radiosonde data for temperature and wind speed, it tends to underestimate seeing at Eltari Airport, Kupang. Despite this discrepancy, the ERA5 data suggest a median seeing of $0.79$ arcseconds at Timau, with optimal seeing conditions in March and December and greater variability during the May to September dry season. These findings are crucial for the planning and operation of the observatory, which requires excellent seeing conditions for its three-band optical imager and a near-infrared camera. Although the seeing at Timau is not as good as some other observatories, the conditions at Timau make it an observatory that has good prospects for equatorial regions.
\end{abstract}

\section{Introduction}
The rapid development of astronomy in recent decades is marked by several episodes of discoveries backed by excellent observing facilities and strategies. An exponential growth of exoplanet discoveries \citep{han2014} is a positive outcome of the deployment of some planet-hunting instruments such as HAT \citep{bakos2004}, SuperWASP \citep{pollacco2006}, Kepler \citep{borucki2010}, and TESS \citep{ricker2015}. Detection of gravitational wave events using extremely precise laser interferometers opened a new era of multi-messenger astronomy \citep{branchesi2018}. Our frontiers in cosmology are expanded by observational results from both ground-based \citep[e.g.,][]{jones2019} and space observatories \citep{goobar2011, aghanim2020}. To extend the frontiers even further, new infrastructures for astronomical observations are under development, while some others have already been commissioned in the past years. In the optical domain, the construction of the 39-m Extremely Large Telescope and its companion instruments shows significant progress \citep{esoreport}. Meanwhile, several medium-sized optical telescopes have started their missions in several countries. New three-meter class telescopes have recently been commissioned and operated in India \citep{sagar2019}, Japan \citep{kurita2020}, Iran \citep{khosroshahi2018}, and Turkey \citep{pirnay2022}.

In order to contribute more seriously to the advancement of astronomy, the Indonesian astronomical community started the construction of a new observatory in the southeastern part of the Indonesian archipelago \citep{mumpuni2018}. Based on multi-year satellite observations, Timau Mountain in Timor Island was identified as a site with the optimum clear sky fraction in the country \citep{hidayat2012}. Recent studies also confirm the general characteristics of the site, placing it as the best among the alternative sites available in the country. Priyatikanto et al. \cite{priyatikanto2023} scrutinized the \emph{in-situ} sky brightness data and estimated the yearly percentage of usable nights at Timau to be 66\%. The available observing time for some parts of the sky, such as regions around the Small Magellanic Cloud, may reach 1200 hours per year. However, the scattering pattern of moonlight observed on-site may indicate an elevated atmospheric extinction \citep{priyatikanto2023}. In line with this finding, Sakti et al. \cite{sakti2023} concluded that the median atmospheric optical depth at Timau is ${\sim}0.13$. The water content above the site is also high, so the atmospheric transparency in the infrared window deteriorates slightly \citep{priyatikanto2024}. Even though Timau serves second-tier sky conditions compared to more ideal sites in mid-latitude regions, promising observation programs at the optical and near-infrared domains can still be conducted.

A 3.8-m telescope with a segmented primary mirror will be the leading facility on site. This telescope is a sister of the Seimei Telescope, which is operated productively at Okayama Observatory, Japan \citep{kurita2020}. A three-band optical imager operated at Pan-STARRRS-like $g$, $r$, and $i$ bands \citep{tonry2012} and a near-infrared camera with $Y$, $J$, and $H$ filters will be the first-generation instruments accompanying the Timau 3.8-m telescope during its first operational years. Afterwards, the next-generation instruments with specialized functionalities are expected to boost the scientific productivity of the telescope even more.

Monitoring the atmospheric environment, managing observations, and selecting astronomical sites require an understanding of seeing parameters \citep{aksaker2020, hellemeier2019}. "Seeing" refers to the angular size of stellar images that are blurred due to atmospheric turbulence. It is defined as the full width at half-maximum of a star image on the focal surface of a large aperture telescope, measured in arcseconds \citep{bi2023}. In relation to seeing, the distribution of wind speed and optical turbulence at astronomical observatories is critical to consider. Optical turbulence is characterized by the refractive index structure parameter $C^2_n$ \citep{tatarskii1961}, which significantly impacts observations made with ground-based telescopes \citep{fried1966, hutt1999}. Atmospheric turbulence inherently limits the angular resolution of optical telescopes and causes motion and blurring of stellar images \citep{roddier1979, vernin1986}.

Wind speed plays a fundamental role in optical astronomy as a key factor influencing optical turbulence. There is a close relationship between wind speed and the intensity of optical turbulence: stronger wind speed gradients are more likely to trigger turbulence. When wind speed increases, the velocity of the turbulent atmospheric layer passing through the telescope's pupil also increases. Furthermore, high wind speeds near the surface can induce vibrations in the telescope structure \citep{sarazin2002}.


These atmospheric parameters and seeing value can then be used to support future planning and development of the Timau site, and should be available so that the temporal variability and associated statistical properties of the Timau site can be perceived.


Recently, the availability of global climate models with adequate spatiotemporal resolutions, such as the latest reanalysis product from the European Center for Medium-Range Weather Forecast (ERA5) allows astronomers to estimate seeing at various locations. To investigate the optical turbulence over the Tibetan Plateau, Han et al. \cite{han2021} computed Richardson's number based on 20-year multi-level ERA5 data. They also demonstrated that the temperature and wind speed profiles from ERA5 are compatible with local radiosonde data. Zhu et al. \cite{zhu2023} explored the statistics of wind speed at 200 hPa pressure level (approximately 12,000 masl) in relation to the seeing at the Tibetan Plateau. At a global scale, a map of astronomical seeing derived from ERA5 data with a $0.25^{\circ}$ resolution is available from the work of Bi et al. \cite{bi2023}. The map shows that sites with excellent seeing conditions can be found in mid-latitude bands. When compared to the \emph{in-situ} results acquired at some sites in China, the median seeing from ERA5 only deviates slightly ($<20\%$). Further investigation on the seeing estimate at other astronomical sites will be of great importance.

This paper presents a deeper analysis of the ERA5 data for evaluating the optical turbulence and seeing statistics Timau National Observatory, Indonesia. Additionally, the availability of radiosonde measurements at Eltari Airport provides additional opportunities to validate the ERA5 data. The data and methods used in this analysis are described in Section \ref{sec:data}. Results and findings are presented in Section \ref{sec:results}. Concluded with a summary in Section \ref{sec:conclusion}, this paper is expected to enlighten the pathway for further development at the Timau National Observatory.

\section{Data and Methods}
\label{sec:data}
\subsection{ERA5 Dataset}
ERA5 provides information on atmospheric properties crucial for estimating seeing. ERA5 is the latest generation of the European Centre for Medium-Range Weather Forecast (ECMWF) reanalysis on global weather and climate \citep{hans2019,hersbach2020}. Pioneered by the FGGE project, completed in 1979 \citep{bengtsson1982}, reanalysis has become an essential part of producing useful atmospheric datasets today. Following that, ERA-15 \citep{era15}, ERA-40 \citep{era40}, and ERA-Interim \citep{era-interim} was published subsequently. ERA5 provides data on atmospheric quantities from 1940 to 5 days behind real-time. This newest dataset has superior qualities compared to ERA-Interim, including hourly output throughout, uncertainty estimate, more atmospheric parameters, and an upgrade on overall quality and level of detail \citep{hans2019}. ERA5 reanalysis combines the atmospheric model with observations from atmospheric stations worldwide to produce a new estimate of global weather quantities with incredible accuracy.

\begin{figure*}
    \centering
    \includegraphics[width=0.9\textwidth]{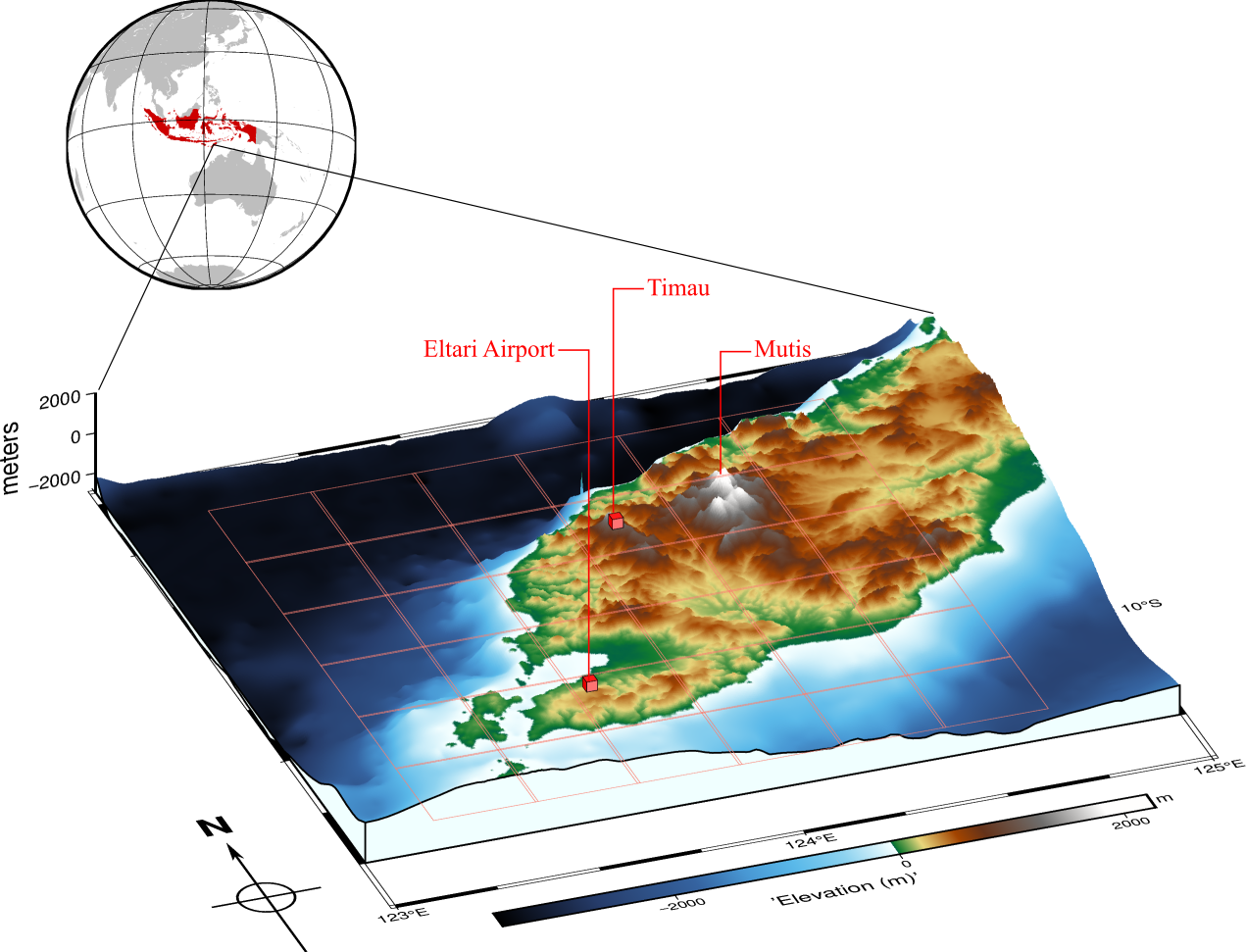}
    \caption{Topography of West Timor where Timau Observatory is situated. The $0.25^{\circ}\times0.25^
    {\circ}$ degree$^2$ grid of ERA5 data is laid over the map.}
    \label{fig:timau}
\end{figure*}

We downloaded the data cube from the Climate Data Store operated under the European Union's Copernicus Program as the input for our analysis. Downloaded as a NetCDF file, the cube mainly contains temperature and horizontal wind speed over Timau and its surroundings (see Figure \ref{fig:timau}) from the surface (1000 hPa) up to 55 km altitude (1 hPa). The horizontal resolution of the data is approximately 30 km, while the interval between pressure levels is 25 hPa. Hourly data from 1 January 2002 to 31 December 2021 were fed into the analysis and the seeing estimation method described in the next section.

Timau ($123.9472^{\circ}$ E, $9.5971^{\circ}$ S) is not located at the centre of the ERA5 rectangular grid. For this reason, we performed multidimensional cubic interpolation to get a proper estimate of the atmospheric parameters over the site. In the process, one-dimensional interpolation is performed consecutively on the deconstructed coordinate system (longitude, latitude, pressure level).

\subsection{Radiosonde at Eltari Airport}
\begin{figure}[b]
    \centering
    \includegraphics[width=\columnwidth]{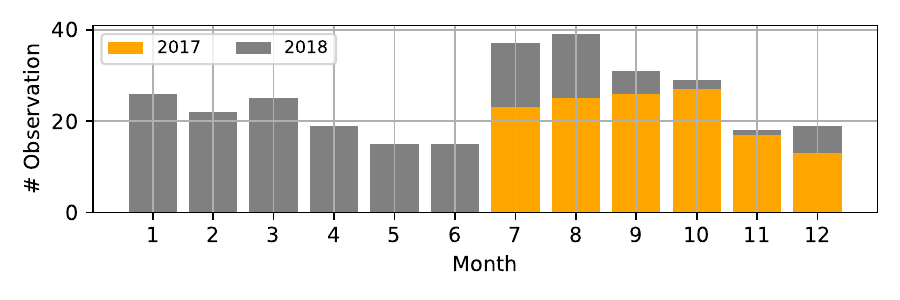}
    \caption{Number of radiosonde measurements at Eltari each month in 2017 and 2018.}
    \label{fig:nradiosonde}
\end{figure}

Since ERA5 is an atmospheric reanalysis model with limited data assimilation, its validity should be evaluated using more reliable \emph{in-situ} measurements. In this regard, we assessed radiosonde data at Eltari Airport ($123.6679^{\circ}$ E, $10.1686^{\circ}$ S) obtained by the Indonesian Agency for Meteorology, Climatology, and Geophysics (BMKG) in 2017-2018. The airport is located approximately 100 km southwest of Timau, which means that the compatibility of ERA5 and radiosonde data at this location reflects the reliability of ERA5 in characterizing seeing at Timau. During the two years mentioned, we identified 295 observations, each conducted around 12:00 UT (approximately 20:00 local time). Figure \ref{fig:nradiosonde} summarizes the number of radiosonde observations each month.

Before thorough analysis, we estimated meteorological parameters (mainly temperature and horizontal wind speed) at different pressure levels, ranging from 10 to 1000 hPa, with 5-hPa intervals. This range covers altitudes up to 37 km above sea level. Linear interpolation was implemented on the ERA5 and radiosonde datasets.

\subsection{Seeing Estimation}
The effect of atmospheric turbulence on astronomical observation is quantified by the seeing parameter $\epsilon$ \citep{roddier1979}:

\begin{equation}
\label{eq:seeing}
    \epsilon = 0.98 \frac{\lambda}{r_0},
\end{equation}

where $\lambda$ is the observing wavelength (e.g., 500 nm for visual) while $r_0$ is the Fried parameter, which defines the effect of inhomogeneity in atmospheric refractive index on the propagation of electromagnetic waves. The Fried parameter is formulated as \citep{fried1966}:
\begin{equation}
\label{eq:friedpar}
    r_0 = \left[ 0.423 \left( \dfrac{2\pi}{\lambda}\right)^2 \int^\infty_0 C_n^2 dh \right]^{-3/5},
\end{equation}
where the refractive index structure, $C_n^2$, is integrated along the line-of-sight.

The refractive index structure constant $C_n^2$ can be associated with several meteorological parameters, as the optical turbulence is dependent on the dynamic condition of the atmosphere \citep{xu2022}. Tatarskii model \citep{tatarskii1961} is commonly used to estimate the $C_n^2$: 
\begin{equation}
    C_n^2 = 2.8 L_0^{4/3} M^2,
\end{equation}
with $L_0$ represents the largest scale of inertial range turbulence while $M$ is the gradient of potential refractive index \citep{Coulman1988}. 

In the context of estimating seeing at the zenith direction, the vertical temperature and the vertical wind speed profiles are fed to the HMNSP99 model \citep{ruggiero2002forecasting}:
\begin{equation}
L_0^{4/3} = 0.1^{4/3} \times 10^{a_1+a_2S-a_3\frac{dT}{dh}}.
\end{equation}
More explicitly, the coefficients $a_1=0.362$, $a_2=16.728$, and $a_3=-192.347$ are applicable for the troposphere, while $a_1=0.757$, $a_2=13.819$, and $a_3=-57.784$ are suitable for the stratosphere. The vertical temperature gradient is denoted by $\frac{dT}{dh}$ whereas wind shears $S$ is:

\begin{equation}
    S= \left[ \left(\frac{\partial u}{\partial h}\right)^2 + \left(\frac{\partial v}{\partial h}\right)^2 \right]^{1/2}
\end{equation}
with $u$ and $v$ represent east-west and north-south horizontal wind components.

Lastly, the gradient of potential refractive index $M$ is computed using the following formula:
\begin{equation}
    M = -7.9 \times 10^{-5}\dfrac{P}{T^2} \dfrac{\partial\theta}{\partial h },
\end{equation}
where $T$ is the atmospheric temperature, $P$ is the atmospheric pressure and $\theta$ is the potential temperature expressed as \citep{xu2022}:
\begin{equation}
\label{eq:potentialtemp}
   \theta = T \left( \frac{100}{P}\right)^{0.286}.
\end{equation}

Formulations above use height in meters, temperature in Kelvin, pressure in millibars (hPa), and speed in meters per second. The resulting seeing parameter ($\epsilon$) is in radians.

\section{Results}
\label{sec:results}

\subsection{Wind Profile Above Eltari}
\begin{figure*}
    \centering
    \includegraphics[width=0.8\textwidth]{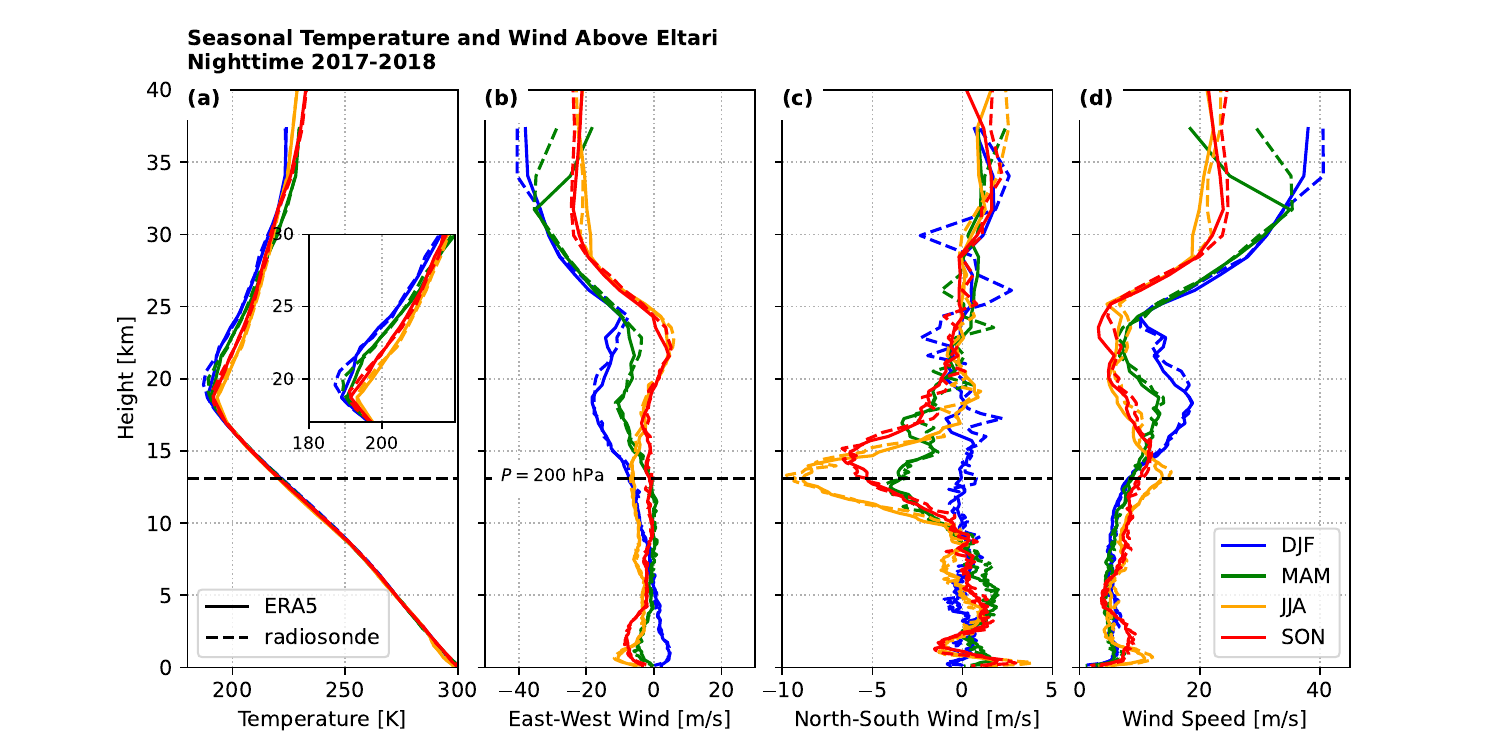}
    \caption{Vertical profile of east-west (a), north-south (b), and horizontal (c) wind speed above Eltari from ERA5 (solid) and radiosonde (dashed) data during four different seasons.}
    \label{fig:seasonalEltari}
\end{figure*}

The compatibility between ERA5 reanalysis data and radiosonde observations was assessed using several key metrics, which are summarized in Table \ref{tab:scores}. For the 2017-2018 period, we found a strong agreement in temperature ($T$) measurements between the two datasets. The coefficient of determination ($R^2$) was nearly perfect at $0.99$, indicating a very high correlation, while the root mean square deviation ($RMSD$) was low at $0.90$ K, suggesting small differences. However, a slight negative bias was also observed, meaning ERA5 temperatures were, on average, slightly lower than those from the radiosondes. A negative bias ($-1.24$ K) in surface temperature wasere also observed when comparing ERA5 data and \emph{in-situ} measurements using a weather station at Eltari Airport \citep{priyatikanto2023}.

Next, the zonal or easterly ($u$) and meridional or northerly ($v$) wind components also showed good agreement. We observed an insignificant bias, meaning there was little systematic difference between the two datasets. The correlation was slightly lower than that for temperature, with a lower $R^2$ value. Meanwhile, the $RMSD$ for wind speed was approximately $2$ m/s.

\begin{table}[b]
    \centering
    \caption{Compatibility scores between atmospheric parameters from ERA5 ($x$) and radiosondes ($x_0$).}
    \label{tab:scores}
    \begin{tabular}{lcrrr}
        \toprule
        Scoring & Formula & \multicolumn{3}{c}{Parameter-wise Score} \\
        & & $u$ [m/s] & $v$ [m/s] & $T$ [K] \\
        \midrule
        $Bias$ & $\langle x - x_0 \rangle$ & $0.05$ & $-0.04$ & $-0.16$ \\
        $R^2$ & $\dfrac{\sum(x-x_0)^2}{\sum\left(x-\langle x\rangle\right)^2}$ & $0.92$ & $0.81$ & $1.00$ \\
        $RMSD$ & $\sqrt{\langle (x-x_0)^2 \rangle}$ & $2.11$ & $2.05$ & $0.90$ \\
        \bottomrule
    \end{tabular}
\end{table}

Vertical profiles of atmospheric temperature below 37 km showed small temporal variation, whereas the vertical wind profiles exhibited clearer seasonal variation.  Figure \ref{fig:seasonalEltari} illustrates the seasonal changes in the zonal (east-west), meridional (north-south), and horizontal wind speeds at various pressure levels. The wind in the troposphere, which is the layer below ${\sim}12$ km, generally had a low speed of less than $10$ m/s. During the wet season (December-January-February), the winds were predominantly eastward. Conversely, during the rest of the year, the winds were mostly westward. This observation is in agreement with the results from Priyatikanto et al. \cite{priyatikanto2024} for the case of wind speed on the ground.

At the 200 hPa pressure level (${\sim}13$ km above sea level), the horizontal wind speed was consistently around 10 m/s for most of the year. However, it was notably stronger during the June-July-August period. A distinct pattern of more prominent southward wind was observed outside of the December-February period. This southward flow appeared to contribute to an increase in wind speed during the dry season, resulting in a maximum horizontal wind speed of around 15 m/s at the specified level. Enhancement in wind speed at 200 hPa is thought to be correlated with a worsening of astronomical seeing conditions \citep{zhu2023, mahrt2013}. However, the wind speed at 13 km above Eltari was significantly lower than the estimated speeds of up to 30 m/s for observatories in mid-latitude regions \citep{hellemeier2019}.

In the stratosphere, the wind profile showed two distinct features: a local peak (or hump) at around 18 km altitude and a general increase in wind speed above 25 km. Seasonal variations were evident in this layer, with higher wind speeds observed during the December-February (DJF) and March-May (MAM) periods. Conversely, lower wind speeds occurred during the June-August (JJA) and September-November (SON) periods. An elevated wind speed at the bottom of the stratosphere, specifically around 20 km, appeared to be another contributing factor to atmospheric turbulence.

Our result was consistent with prior research. For instance, Szkolka et al. \cite{szkolka2025} reported a strong agreement between horizontal wind data from ERA5 and observations using Equatorial Atmospheric Radar \citep[EAR,][]{fukao2003} in West Sumatra. However, a significant discrepancy was noted in the vertical wind data. The bias in the ERA-5 reanalysis for vertical wind and its diurnal variability is likely due to the model's inability to accurately represent the evolution of atmospheric convection over the complex coastal topography. This topographical influence is further supported by the daily mean profile's bias, as neighboring coastal grid points show more detail and better alignment with EAR observations. Fortunately, while vertical wind speed is crucial for cloud formation, its impact on turbulence and seeing is minimal.

\subsection{Seeing at Eltari}
\begin{figure*}
    \centering
    \includegraphics[width=0.7\textwidth]{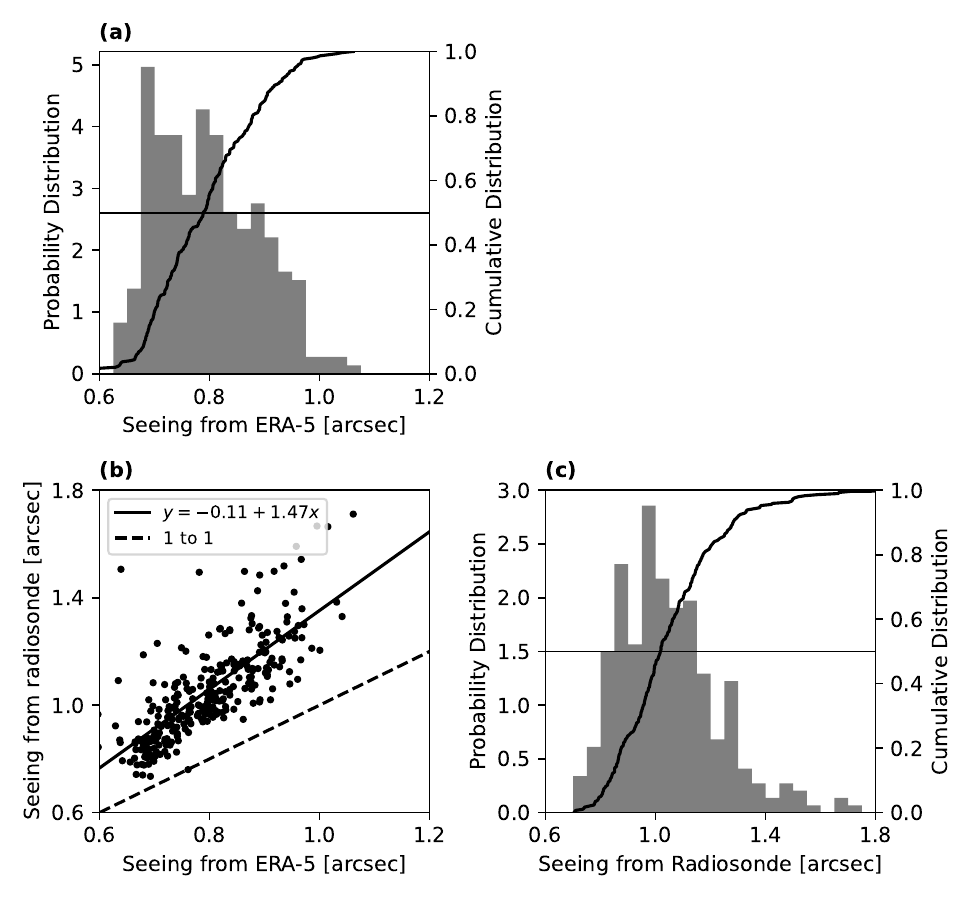}
    \caption{Comparison between seeing from ERA5 and radiosonde.}
    \label{fig:eltariSeeing}
\end{figure*}

\begin{table*}
    \centering
    \caption{Monthly and global aggregate of seeing parameter (in arcseconds) at Timau and Eltari.}
    \label{tab:summary}
    \begin{tabular}{lrrrrrrrrr}
    \toprule
    Month & \multicolumn{3}{c}{Timau-ERA5} & \multicolumn{3}{c}{Eltari-ERA5} & \multicolumn{3}{c}{Eltari-radiosonde} \\
     & $Q_1$ & $Q_2$ & $Q_3$ & $Q_1$ & $Q_2$ & $Q_3$ & $Q_1$ & $Q_2$ & $Q_3$ \\
    \midrule
    Jan & 0.77 & 0.79 & 0.81 & 0.69 & 0.71 & 0.73 & 0.83 & 0.88 & 0.97 \\
    Feb & 0.77 & 0.79 & 0.81 & 0.70 & 0.73 & 0.76 & 0.86 & 0.89 & 0.96 \\
    Mar & 0.76 & 0.78 & 0.79 & 0.69 & 0.76 & 0.79 & 0.87 & 0.95 & 1.00 \\
    Apr & 0.77 & 0.78 & 0.80 & 0.75 & 0.79 & 0.83 & 0.94 & 0.99 & 1.06 \\
    May & 0.78 & 0.80 & 0.83 & 0.70 & 0.78 & 0.84 & 0.85 & 0.98 & 1.08 \\
    Jun & 0.80 & 0.83 & 0.87 & 0.79 & 0.87 & 0.90 & 1.05 & 1.13 & 1.22 \\
    Jul & 0.81 & 0.85 & 0.90 & 0.86 & 0.90 & 0.96 & 1.08 & 1.17 & 1.29 \\
    Aug & 0.81 & 0.85 & 0.91 & 0.83 & 0.88 & 0.93 & 1.05 & 1.16 & 1.27 \\
    Sep & 0.78 & 0.81 & 0.84 & 0.81 & 0.84 & 0.87 & 1.05 & 1.13 & 1.18 \\
    Oct & 0.75 & 0.77 & 0.80 & 0.75 & 0.79 & 0.81 & 0.97 & 1.02 & 1.10 \\
    Nov & 0.74 & 0.76 & 0.78 & 0.68 & 0.70 & 0.71 & 0.81 & 0.85 & 0.92 \\
    Dec & 0.75 & 0.77 & 0.79 & 0.69 & 0.70 & 0.74 & 0.84 & 0.88 & 0.97 \\
    \midrule
    Overall & 0.77 & 0.79 & 0.82 & 0.72 & 0.79 & 0.86 & 0.92 & 1.02 & 1.15 \\
    Obs. season & 0.78 & 0.81 & 0.85 & 0.79 & 0.84 & 0.90 & 1.01 & 1.10 & 1.21 \\
    \bottomrule
    \end{tabular}
\end{table*}

Based on the vertical profiles of atmospheric temperature and wind speed, we calculated the astronomical seeing using the formulas detailed in Section \ref{sec:data}. The results (summarized in Table \ref{tab:summary}) showed a systematic difference between the seeing values derived from ERA5 reanalysis data and those from radiosonde observations. Meanwhile, ERA5-derived seeing values typically ranged from $0.60$ to $1.10$ arcseconds, with a median of $0.79$ arcseconds. In contrast, the seeing values calculated from radiosonde data had a higher median of $1.01$ arcseconds and a wider range, extending up to $1.80$ arcseconds. As shown in  Figure \ref{fig:eltariSeeing}, the ERA5 data consistently underestimated the seeing compared to the radiosonde data. Linear fit to the data yields a slope of $1.47$ with a coefficient of determination $R^2=0.51$.

Research shows a difference between seeing values calculated using ERA5 and those derived from other methods, such as radiosonde data or Differential Image Motion Monitor \citep[DIMM,][]{sarazin1990, tokovinin2002} techniques. Bi et al. \cite{bi2023} compared seeing estimates at seven Chinese sites and found significant variations. For example, at the Ali site, ERA5-derived seeing values were only 80\% of those obtained from optical observations, representing the most significant underestimation. In contrast, ERA5 values at the Lenghu site were 128\% of the measured values. Meanwhile, our study found that ERA5 estimates were approximately 76\% of the values calculated using radiosonde data.

\begin{figure*}[t]
    \centering
    \includegraphics[width=0.8\textwidth]{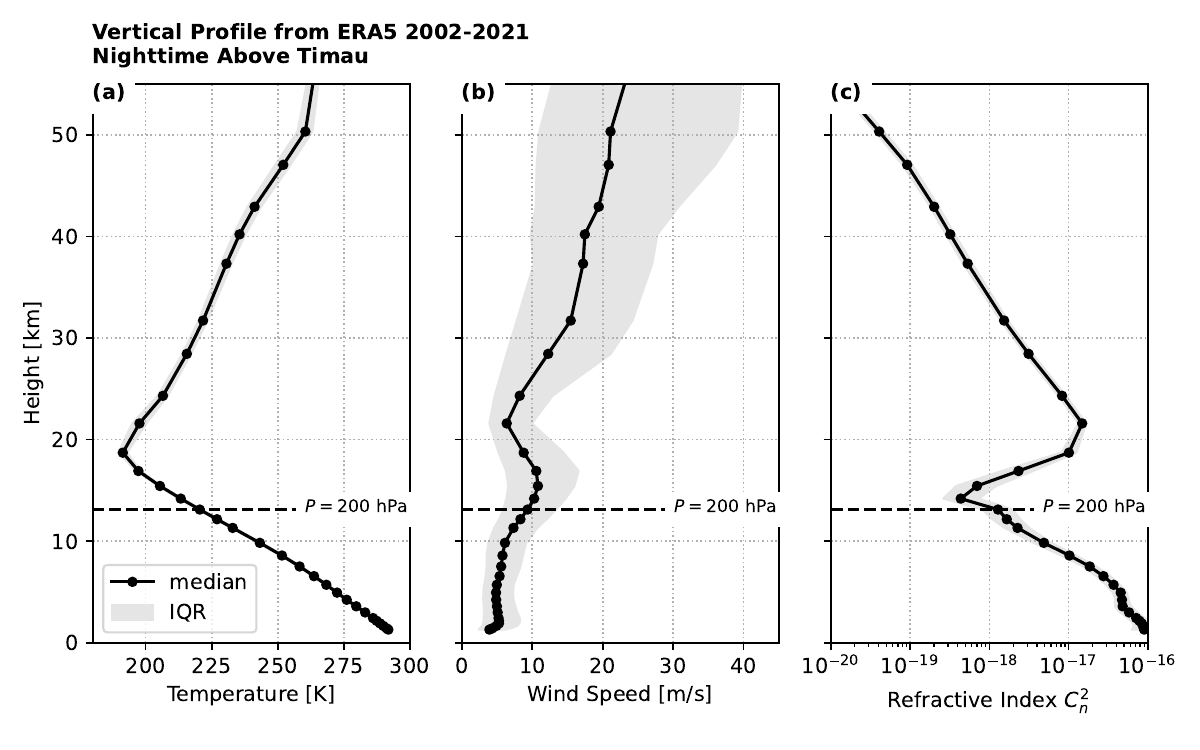}
    \caption{Atmospheric profile over Timau as indicated by temperature (a), horizontal wind speed (b), and refractive index structure constant (c). The median and the inter-quartile range (IQR) are from 2002-2021 aggregated at different pressure levels.}
    \label{fig:profileTimau}
\end{figure*}

\begin{figure*}
    \centering
    \includegraphics[width=0.8\textwidth]{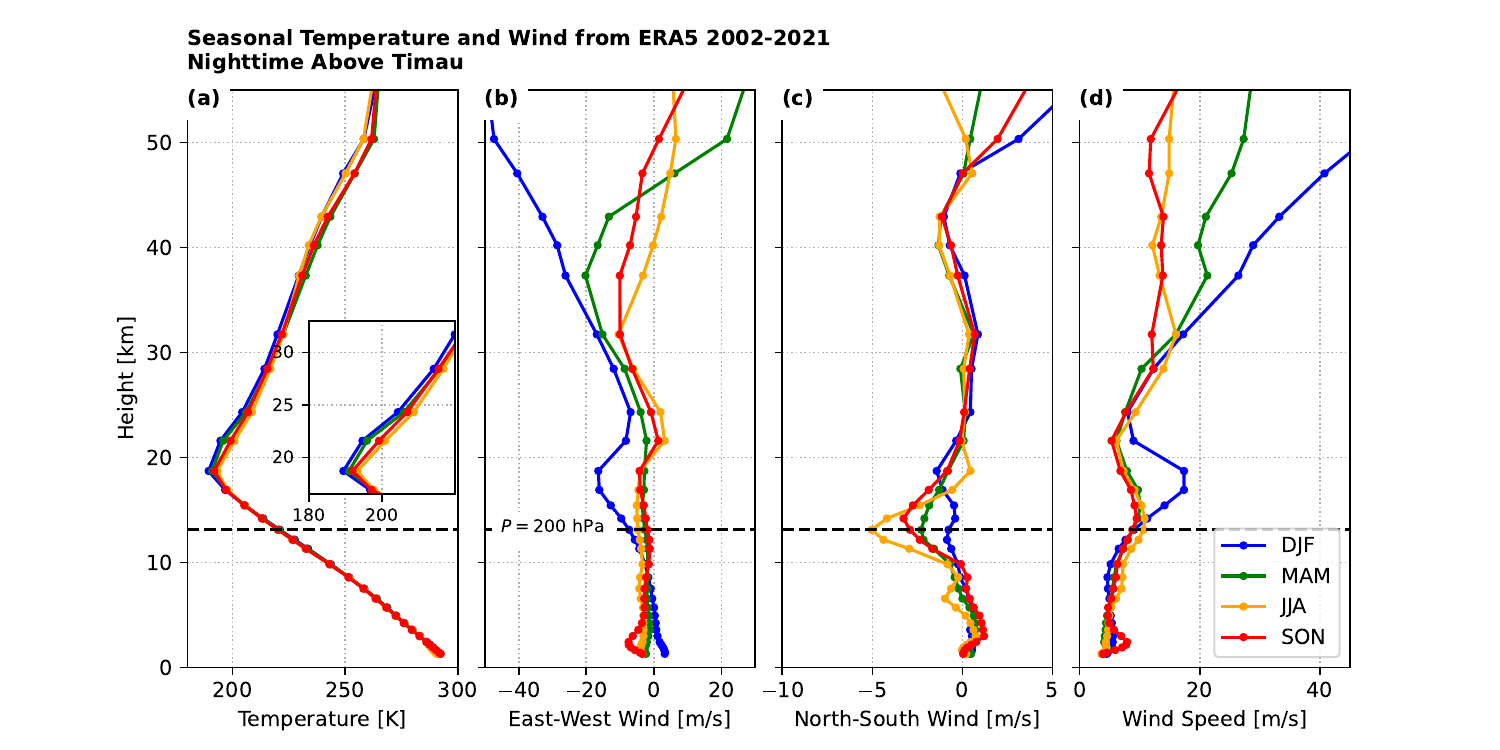}
    \caption{Seasonal variation of wind profile over Timau.}
    \label{fig:seasonalTimau}
\end{figure*}

The discrepancy in seeing values between ERA5 and radiosonde data can be attributed to the differences in their wind speed profiles. As shown in Figure \ref{fig:seasonalEltari}, the radiosonde data exhibit significantly greater fluctuations in wind speed compared to the ERA5 data. Astronomical seeing is directly linked to atmospheric turbulence, which is generated by wind speed gradients. A higher gradient--meaning a sharper change in wind speed with altitude--leads to more turbulence and, consequently, worse seeing. The radiosonde data, with larger fluctuations, indicate a more turbulent atmosphere than what is represented by the ERA5 model. Moreover, ERA5 data present a less-detailed, smoother wind profile; it systematically underestimates the actual atmospheric turbulence and thus, the resulting seeing. The radiosonde, by capturing these finer-scale fluctuations, provides a more accurate representation of the conditions that contribute to turbulence.

\subsection{Atmospheric Profile Above Timau}
The atmospheric profiles on Timau are shown in Figure \ref{fig:profileTimau}, while the seasonal profiles for wind speed are shown in Figure \ref{fig:seasonalTimau}.

On average, the surface temperature at Timau site is approximately 300 K. Temperature then steadily decreases throughout the troposphere, a stable trend that continues up to an altitude of nearly 20 km. At this point, the temperature decrease halts and begins to increase with altitude, marking the transition into the stratosphere. ERA5 data revealed that the equatorial region has a thicker troposphere compared to mid-latitude regions, with the temperature inversion in the stratosphere occurring at a higher altitude Han et al. \cite{han2021}. Furthermore, the data indicate that temperature variation remains relatively small throughout the year.

Turbulence is closely related to temperature and wind speed, among other things. A change in temperature trend (from decreasing to increasing) occurs at an altitude of around 20 km, where a sharp temperature increase occurs \citep{wallace2006}. However, this increase is not as sharp as in the lower layers, making it more stable. Under these conditions, turbulence occurs more frequently in the lower atmosphere, where warm temperatures also contribute to its formation.

In contrast to the stable temperature profile, the stratospheric wind speed profile shows greater variability. This, combined with the higher temperatures in this layer, results in a larger inertial range turbulence ($L_0$) in the stratosphere compared to other layers \citep{xu2022}.  Figure \ref{fig:seasonalTimau} presents the seasonal changes of wind speed profiles. As illustrated in this figure, the 200 hPa pressure level is a critical region in the stratosphere where meridional (north-south) wind exhibits significant seasonal changes. These changes are analogous to those observed in the Eltari data (Figure \ref{fig:seasonalEltari}. Specifically, southward winds reach their maximum strength during June-August. Meanwhile, variations in zonal (east-west) wind are most pronounced at an altitude of approximately 18 km. At this height, the westward wind is at its strongest during the December-February (DJF) period.

Refractive index structure ($C_n^2$) in the context of the atmosphere and astronomical observations in the optical window usually refers to the vertical and spatial variations of the air's refractive index, which trigger atmospheric turbulence, affect light propagation, and reduce seeing condition. Indicated in the rightmost panel of Figure \ref{fig:profileTimau}, the vertical profile of $C_n^2$ shows that the refractive index value is most significant at the surface, which is generally caused by convection and wind shear \citep{fairall1984}. Then it decreases with increasing altitude, and at a certain altitude (about 15 km, which is the layer between the troposphere and the stratosphere) it gets larger again, and decreases again in the stratosphere. Considering that the seeing parameter is a turbulence integral along the line of sight, we can say that more than half of the seeing is contributed by surface turbulence in the lower atmosphere.

\subsection{Seeing at Timau}
\begin{figure}
    \centering
    \includegraphics[width=0.45\textwidth]{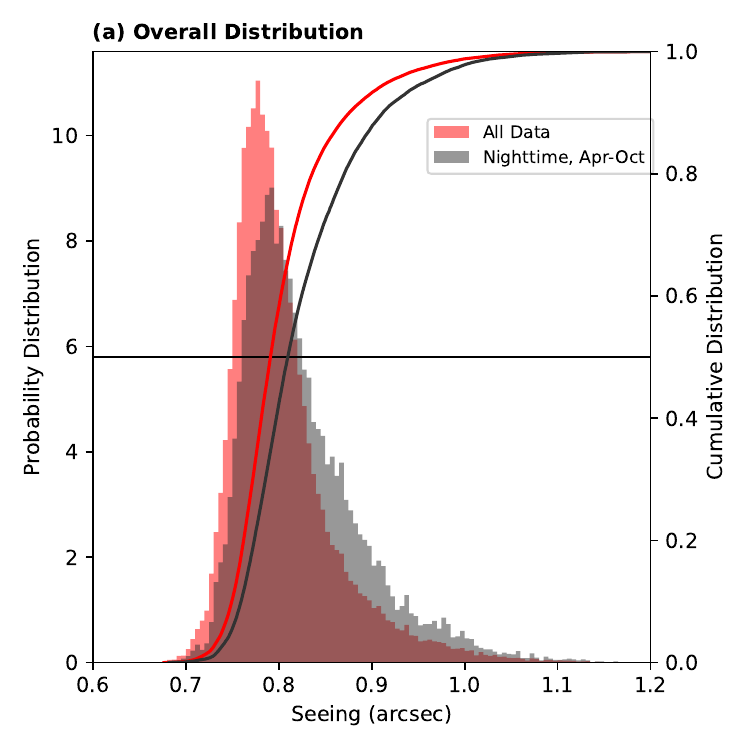}
    \includegraphics[width=0.45\textwidth]{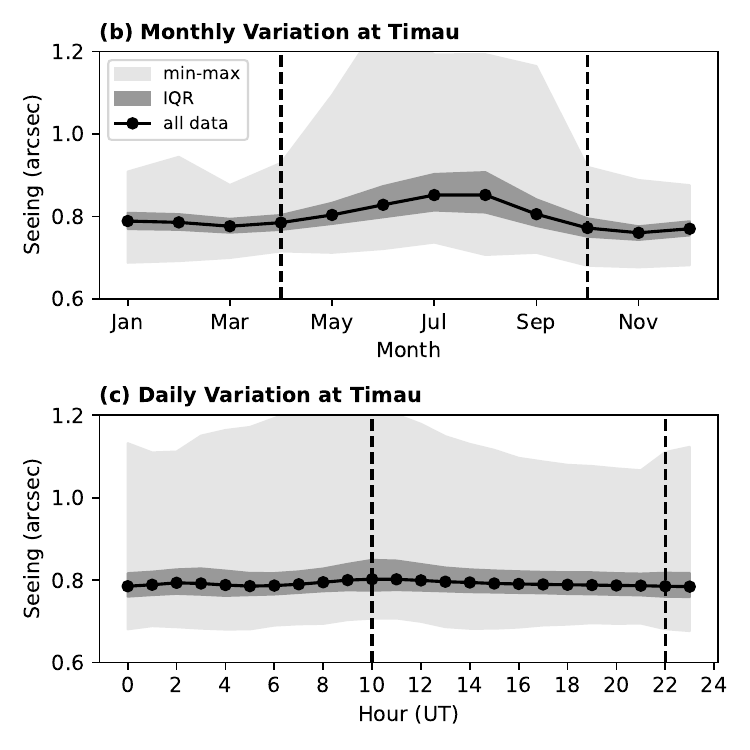}
    \caption{Overall distribution (a), seasonal change (b), and daily variation (c) of seeing at Timau.}
    \label{fig:seeingTimau}
\end{figure}

Seeing estimates based on the ERA5 datasets (from 2002 to 2021) are summarized in Figure \ref{fig:seeingTimau} and Table \ref{tab:summary}. The median seeing value was found to be $0.79$ arcseconds, which was similar to that of Eltari. In more detail, the 95th percentile was $0.92$ arcseconds, indicating that the seeing was better than $0.92$ arcseconds for 95\% of the time.

Our analysis of monthly and daily seeing values revealed some other findings. Firstly, the best seeing conditions (lowest values) were observed during March and November, also with minimal variability. In contrast, from May to September, a period that corresponds to the dry season, the seeing values were consistently higher (worse) and showed greater variability ($Q_1=0.78, Q_2=0.81, Q_3=0.85$). While the dry season is traditionally considered the prime observing period, a closer look at the data indicates that specific months outside this period, such as March and November, may offer superior seeing conditions.

Secondly, the daily profile showed that seeing was at its worst (highest value) around 10:00 UT (18:00 local time) when the Sun is setting and the local temperature changes significantly. The seeing then steadily improved and stabilized throughout the night, which is favorable for astronomical observations.

The findings above align closely with prior direct observations. For instance, in July and August 2019, Akbar et al. \cite{akbar2019} used the DIMM \citep{sarazin1990, tokovinin2002} technique at Timau and recorded a median seeing of 0.93 arcseconds. Another study by Saputra et al. \cite{saputra2022} in 2021 also arrived at a similar conclusion. Assuming the underestimation factor from the Eltari radiosonde analysis holds true for Timau, the ERA5-based median seeing of 0.79 arcseconds should be multiplied by 1.3, resulting in a corrected value of 1.03 arcseconds. This adjusted figure aligns more closely with direct measurements and can serve as a valuable reference for future observatory planning and operations.

The minor discrepancies between our ERA5 estimates and the DIMM measurements are likely due to a couple of factors. First, the ERA5 data may not fully capture the significant fluctuations in wind speeds at low altitudes. Additionally,
Timau is located at an altitude of approximately 1300 meters above sea level and is situated on a modest plain surrounded by a diverse topography of hills and valleys (see Figure \ref{fig:timau}. This varied topography, as indicated by Li et al. \cite{li2022}, can lead to local wind variations and turbulence, which naturally degrade seeing conditions.

The peak of Timau is not the highest in the area. The highest peak is Mount Mutis, located 35 km East of Timau, with an altitude of approximately 2,400 meters above sea level. During dry season, the surface wind blows from East to West. The percentage of westerly winds with speeds greater than 10 m/s reaches almost 40\% in June, July, and August \citep{priyatikanto2024}. While passing Mount Mutis, the winds may introduce additional turbulence that reduces seeing conditions at Timau.

The complex topography of the region, particularly toward Mutis, likely contributes to atmospheric turbulence. As confirmed by studies like Bowen et al. \cite{bowen1977} and Gawad et al. \cite{gawad2006}, these sortkind of topographical features can alter wind speed and direction, which increases the likelihood of turbulence. During the dry season, westerly winds with speeds exceeding 10 m/s are more common, leading to deteriorated seeing during this time. On the other hand, the observatory's proximity to the sea—approximately 25 km to the west—also influences local conditions. This closeness can increase atmospheric water vapor and lead to rapid temperature fluctuations. However, Priyatikanto et al. \cite{priyatikanto2024} noted that winds from the west (and thus from the sea) primarily occur in December, January, and February.

\section{Reflection}
The Timau National Observatory features a 3.8-meter optical telescope as its primary instrument, which is the largest optical telescope in Southeast Asia. The telescope will initially be equipped with two first-generation instruments. First, \textsc{trioptika} is an imaging system designed for optimal performance under Timau's atmospheric conditions, with a nominal seeing of 1 arcsecond. The camera operates at an $f/2$ focal ratio, providing a $12.5 \times 12.5$ square-arcminute field of view and a pixel scale of $0.37$ arcseconds. Second, \textsc{nirka} has an infrared sensor with a pixel scale almost similar to the optical imager. Both imagers meet the Nyquist sampling criteria, ensuring that the spatial information from the telescope's optics is adequately captured given the prevailing atmospheric seeing conditions. This makes them well-suited for high-quality imaging and astrometric research.

In addition to imaging, the 3.8-meter telescope will also host a low-resolution spectrograph. This instrument is being developed through an international collaboration with the University of Science and Technology of China, the University of Turku (Finland), and Kyoto University (Japan), and is specifically intended for supernova observation. For spectroscopy, the impact of atmospheric seeing is less critical as long as the target object is centered within the spectrograph's aperture. Furthermore, if an integral field spectrograph is planned for the future, the site's atmospheric seeing data will be essential for optimizing its performance.

The observatory will also support various surveys using smaller-aperture telescopes. An example is the MASTER (Mobile Astronomical System of TElescope-Robots) network \citep{kornilov2012}, which employs a 40 cm telescope to observe transient phenomena \citep{lipunov2022}. For this mission, atmospheric seeing is not a limiting factor because the seeing value is smaller than the instrument's 1.85 arcsecond pixel scale, making the MASTER telescope's observations robust to atmospheric blurring.

\section{Conclusion}
\label{sec:conclusion}

This paper mainly focuses on the long-term atmospheric parameters at Timau National Observatory in Indonesia based on the ERA5 dataset, from 2002 to 2021. Aiming to estimate the seeing characteristics at Timau, it comes with the following key findings:
\begin{itemize}
    \item Analysis of radiosonde data from Eltari Airport (approximately 100 km southwest of Timau) proves the compatibility between ERA5 and direct measurements of the key meteorological parameters. However, ERA5's smoother wind profiles may underestimate (by a factor of $1.47$) actual atmospheric turbulence compared to radiosonde data.
    \item Despite the discrepancies in seeing values, the ERA5 dataset shows good agreement with radiosonde data for temperature ($RMSD=0.48$ K, $bias=-0.16$ K) and horizontal wind speed ($RMSD=2.20$ m/s, $bias=-0.54$ m/s).
    \item At Timau, the median seeing is $0.79$ arcsec, with the 95th percentile at $0.92$ arcsec. This finding complements the prior knowledge that was established from direct seeing measurements conducted in a handful of observing campaigns.
    \item The best seeing conditions are observed around March and December, while greater variability occurs between May and September (the dry season).
\end{itemize}

\section{Acknowledgments}
We appreciate the courtesy of the Eltari Meteorological Station (BMKG) for the provision of radiosonde data utilized in this paper. We thank Bambang Suhandi, who made this data available to us.

This preprint has not undergone peer review or any post-submission improvements or corrections. The Version of Record of this article is published in Experimental Astronomy, and is available online at https://doi.org/10.1007/s10686-026-10054-y.

\bibliography{main}

@article{branchesi2018,
    title={GW170817: The Dawn of Multi-messenger Astronomy Including Gravitational Waves},
    author={Branchesi, Marica},
    journal={Multiple Messengers and Challenges in Astroparticle Physics},
    pages={489--497},
    year={2018},
    publisher={Springer}
}

@article{pollacco2006,
    title={The WASP project and the SuperWASP cameras},
    author={Pollacco, Don L and Skillen, I and Cameron, A Collier and Christian, DJ and Hellier, C and Irwin, J and Lister, TA and Street, RA and West, Richard G and Anderson, D and others},
    journal={Publications of the Astronomical Society of the Pacific},
    volume={118},
    number={848},
    pages={1407},
    year={2006},
    publisher={IOP Publishing}
}

@article{ricker2015,
    title={Transiting exoplanet survey satellite},
    author={Ricker, George R and Winn, Joshua N and Vanderspek, Roland and Latham, David W and Bakos, G{\'a}sp{\'a}r {\'A} and Bean, Jacob L and Berta-Thompson, Zachory K and Brown, Timothy M and Buchhave, Lars and Butler, Nathaniel R and others},
    journal={Journal of Astronomical Telescopes, Instruments, and Systems},
    volume={1},
    number={1},
    pages={014003--014003},
    year={2015},
    publisher={Society of Photo-Optical Instrumentation Engineers}
}

@article{han2014,
    title={Exoplanet orbit database. II. Updates to exoplanets. org},
    author={Han, Eunkyu and Wang, Sharon X and Wright, Jason T and Feng, Y Katherina and Zhao, Ming and Fakhouri, Onsi and Brown, Jacob I and Hancock, Colin},
    journal={Publications of the Astronomical Society of the Pacific},
    volume={126},
    number={943},
    pages={827},
    year={2014},
    publisher={IOP Publishing}
}

@article{bakos2004,
    title={Wide-Field Millimagnitude Photometry with the HAT: A Tool for Extrasolar Planet Detection},
    author={Bakos, Gaspar and Noyes, RW and Kov{\'a}cs, G and Stanek, KZ and Sasselov, DD and Domsa, Istv{\'a}n},
    journal={Publications of the Astronomical Society of the Pacific},
    volume={116},
    number={817},
    pages={266},
    year={2004},
    publisher={IOP Publishing}
}

@article{borucki2010,
    title={Kepler planet-detection mission: introduction and first results},
    author={Borucki, William J and Koch, David and Basri, Gibor and Batalha, Natalie and Brown, Timothy and Caldwell, Douglas and Caldwell, John and Christensen-Dalsgaard, J{\o}rgen and Cochran, William D and DeVore, Edna and others},
    journal={Science},
    volume={327},
    number={5968},
    pages={977--980},
    year={2010},
    publisher={American Association for the Advancement of Science}
}

@article{aghanim2020,
    title={Planck 2018 results-I. Overview and the cosmological legacy of Planck},
    author={Aghanim, Nabila and Akrami, Yashar and Arroja, Frederico and Ashdown, Mark and Aumont, J and Baccigalupi, Carlo and Ballardini, M and Banday, Anthony J and Barreiro, RB and Bartolo, Nicola and others},
    journal={Astronomy \& Astrophysics},
    volume={641},
    pages={A1},
    year={2020},
    publisher={EDP sciences}
}

@article{goobar2011,
    title={Supernova cosmology: legacy and future},
    author={Goobar, Ariel and Leibundgut, Bruno},
    journal={Annual Review of Nuclear and Particle Science},
    volume={61},
    pages={251--279},
    year={2011},
    publisher={Annual Reviews}
}

@article{aksaker2020,
    title={Global site selection for astronomy},
    author={Aksaker, Nazim and Yerli, Sinan Kaan and Erdo{\u{g}}an, Ma and Kurt, Z and Kaba, K and Bayazit, M and Yesilyaprak, C},
    journal={Monthly Notices of the Royal Astronomical Society},
    volume={493},
    number={1},
    pages={1204--1216},
    year={2020},
    publisher={Oxford University Press}
}

@article{hidayat2012,
    title={Clear sky fraction above Indonesia: an analysis for astronomical site selection},
    author={Hidayat, T and Mahasena, P and Dermawan, B and Hadi, TW and Premadi, PW and Herdiwijaya, D},
    journal={Monthly Notices of the Royal Astronomical Society},
    volume={427},
    number={3},
    pages={1903--1917},
    year={2012},
    publisher={Blackwell Science Ltd Oxford, UK}
}

@article{priyatikanto2023,
    title={Characterization of Timau National Observatory using limited in situ measurements},
    author={Priyatikanto, Rhorom and Mumpuni, Emanuel Sungging and Hidayat, Taufiq and Saputra, Muhamad Bayu and Murti, Mulya Diana and Rachman, Abdul and Yatini, Clara Yono},
    journal={Monthly Notices of the Royal Astronomical Society},
    volume={518},
    number={3},
    pages={4073--4083},
    year={2023},
    publisher={Oxford University Press}
}

@article{priyatikanto2024,
    title={Weather conditions at Timau National Observatory from ERA5},
    author={Priyatikanto, Rhorom and Admiranto, Agustinus Gunawan and Djamaluddin, Thomas and Rachman, Abdul and Wijaya, Dudy D.},
    journal={PASA},
    volume={xx},
    number={xx},
    pages={xx},
    year={2024},
    publisher={Cambridge University Press}
}

@article{mumpuni2018,
    title={Future astronomy facilities in Indonesia},
    author={Mumpuni, Emanuel Sungging and Puspitarini, Lucky and Priyatikanto, Rhorom and Yatini, Clara Y and Putra, Mahasena},
    journal={Nature Astronomy},
    volume={2},
    number={12},
    pages={930--932},
    year={2018},
    publisher={Nature Publishing Group UK London}
}

@article{hersbach2020,
    title={The ERA5 global reanalysis},
    author={Hersbach, Hans and Bell, Bill and Berrisford, Paul and Hirahara, Shoji and Hor{\'a}nyi, Andr{\'a}s and Mu{\~n}oz-Sabater, Joaqu{\''\i}n and Nicolas, Julien and Peubey, Carole and Radu, Raluca and Schepers, Dinand and others},
    journal={Quarterly Journal of the Royal Meteorological Society},
    volume={146},
    number={730},
    pages={1999--2049},
    year={2020},
    publisher={Wiley Online Library}
}

@article{jones2019,
    title={The foundation supernova survey: measuring cosmological parameters with supernovae from a single telescope},
    author={Jones, DO and Scolnic, DM and Foley, RJ and Rest, A and Kessler, R and Challis, PM and Chambers, KC and Coulter, DA and Dettman, KG and Foley, MM and others},
    journal={The Astrophysical Journal},
    volume={881},
    number={1},
    pages={19},
    year={2019},
    publisher={American Astronomical Society}
}

@article{khosroshahi2018,
    title={Linking a noble past to future challenges},
    author={Khosroshahi, Habib G},
    journal={Nature Astronomy},
    volume={2},
    number={5},
    pages={429--429},
    year={2018},
    publisher={Nature Publishing Group UK London}
}

@article{sarazin1990,
    title={The ESO differential image motion monitor},
    author={Sarazin, Marc and Roddier, F},
    journal={Astronomy and Astrophysics},
    volume={227},
    pages={294--300},
    year={1990}
}

@article{tokovinin2002,
    title={From differential image motion to seeing},
    author={Tokovinin, Andrei},
    journal={Publications of the Astronomical Society of the Pacific},
    volume={114},
    number={800},
    pages={1156},
    year={2002},
    publisher={IOP Publishing}
}

@article{akbar2019,
    title={Report on Sky Brightness, Seeing, and Weather Measurements at Timau Observatory, East Nusa Tenggara},
    author={Akbar, Evan Irawan and Jatmiko, Agus Triyono Puri and Putra, Mahasena and Nurzaman, Muhammad Zamzam and Mumpuni, Emanuel Sungging and Mandey, Denny and Raharto, Moedji},
    journal={Journal of Physics: Conference Series},
    volume={1245},
    number={1},
    pages={012024},
    year={2019},
    publisher={IOP Publishing}
}

@article{bi2023,
    title={Investigation of the Global Spatio-Temporal Characteristics of Astronomical Seeing},
    author={ Bi, Cuicui and Qing, Chun and Qian, Xianmei and Luo, Tao and Zhu, Wenyue and Weng, Ningquan},
    journal={Remote Sensing},
    volume={15},
    number={9},
    pages={2225},
    year={2023},
    publisher={MDPI}
}

@article{saputra2022,
    title={Report on seeing, sky brightness, and meteorological properties measurements at Timau National Observatory site},
    author={Saputra, MB and Danarianto, MD and Murti, MD and Alwan, MA and Yanti, RJ and others},
    journal={Journal of Physics: Conference Series},
    volume={2214},
    number={1},
    pages={012013},
    year={2022},
    publisher={IOP Publishing}
}

@techreport{esoreport,
    title={Annual Report 2022},
    author={Barcons, Xavier},
    institution={European Southern Observatory},
    year={2022}
}

@article{sagar2019,
    title={The 3.6 metre Devasthal Optical Telescope},
    author={Sagar, Ram and Kumar, Brijesh and Omar, Amitesh},
    journal={Current Science},
    volume={117},
    number={3},
    pages={365--381},
    year={2019},
    publisher={JSTOR}
}

@article{sakti2023,
    title={Machine learning-based spatial data development for optimizing astronomical observatory sites in Indonesia},
    author={Sakti, Anjar Dimara and Zakiar, Muhammad Rizky and Santoso, Cokro and Windasari, Nila Armelia and Jaelani, Anton Timur and Damayanti, Seny and Anggraini, Tania Septi and Putri, Anissa Dicky and Hudalah, Delik and Deliar, Albertus},
    journal={Plos one},
    volume={18},
    number={10},
    pages={e0293190},
    year={2023},
    publisher={Public Library of Science San Francisco, CA USA}
}

@article{kurita2020,
    title={The Seimei telescope project and technical developments},
    author={Kurita, Mikio and Kino, Masaru and Iwamuro, Fumihide and Ohta, Kouji and Nogami, Daisaku and Izumiura, Hideyuki and Yoshida, Michitoshi and Matsubayashi, Kazuya and Kuroda, Daisuke and Nakatani, Yoshikazu and others},
    journal={Publications of the Astronomical Society of Japan},
    volume={72},
    number={3},
    pages={48},
    year={2020},
    publisher={Oxford University Press}
}

@article{zhu2023,
    title={Astronomical seeing and wind speed distributions with ERA5 data at Lenghu site on the Tibetan Plateau},
    author={Zhu, Liming and Zhang, Hanjiu and Sun, Gang and Li, Xuebin and Yang, Fan and He, Fei and Weng, Ningquan and Deng, Licai},
    journal={Monthly Notices of the Royal Astronomical Society},
    volume={522},
    number={1},
    pages={1419--1427},
    year={2023},
    publisher={Oxford University Press}
}

@article{han2021,
    title={Analysis of wind-speed profiles and optical turbulence above Gaomeigu and the Tibetan Plateau using ERA5 data},
    author={Han, Yajuan and Yang, Qike and Liu, Nana and Zhang, Kun and Qing, Chun and Li, Xuebin and Wu, Xiaoqing and Luo, Tao},
    journal={Monthly Notices of the Royal Astronomical Society},
    volume={501},
    number={4},
    pages={4692--4702},
    year={2021},
    publisher={Oxford University Press}
}

@inproceedings{pirnay2022,
    title={DAG 4m telescope: optics completion, on-site integration and test},
    author={Pirnay, Olivier and Albart, Pierre and Bastin, Christian and De Ville, Jonathan and Gabriel, Eric and Leseur, Thibault and Lousberg, Gr{\'e}gory P and M{\'e}ant, Laurence and Orban, Sabrina and Tortolani, Jean-Marc and others},
    booktitle={Ground-based and Airborne Telescopes IX},
    volume={12182},
    pages={1327--1338},
    year={2022},
    organization={SPIE}
}

@article{hans2019,
    author = {Hans Hersbach and W Bell and P. Berrisford and Andras Horányi and Muñoz-Sabater J. and J. Nicolas and Raluca Radu and Dinand Schepers and Adrian Simmons and Cornel Soci and Dick Dee},
    title = {Global reanalysis: goodbye ERA-Interim, hello ERA5},
    year = {2019},
    journal = {ECMWF Newsletter},
    chapter = {Meteorology},
    pages = {17-24},
    month = {04/2019},
    language = {eng}
}

@article {bengtsson1982,
    author = "L. Bengtsson and M. Kanamitsu and P. Kållberg and S. Uppala",
    title = "FGGE Research Activities at ECMWF",
    journal = "Bulletin of the American Meteorological Society",
    year = "1982",
    publisher = "American Meteorological Society",
    address = "Boston MA, USA",
    volume = "63",
    number = "3",
    pages=      "277 - 303"
}

@article{lipunov2022,
  title={MASTER real-time multi-message observations of high energy phenomena},
  author={Lipunov, Vladimir M and Kornilov, Viktor G and Zhirkov, Kirill and Kuznetsov, Artem and Gorbovskoy, Evgenii and Budnev, Nikolai M and Buckley, David AH and Lopez, Rafael Rebolo and Serra-Ricart, Miquel and Francile, Carlos and others},
  journal={Universe},
  volume={8},
  number={5},
  pages={271},
  year={2022},
  publisher={MDPI}
}

@article{kornilov2012,
  title={Robotic optical telescopes global network MASTER II. Equipment, structure, algorithms},
  author={Kornilov, Victor G and Lipunov, Vladimir M and Gorbovskoy, Evgeny S and Belinski, Aleksander A and Kuvshinov, Dmitry A and Tyurina, Natalia V and Shatsky, Nikolai I and Sankovich, Anatoly V and Krylov, Aleksander V and Balanutsa, Pavel V and others},
  journal={Experimental Astronomy},
  volume={33},
  number={1},
  pages={173--196},
  year={2012},
  publisher={Springer}
}

@article{era15,
    title={ERA description},
    author={Gibson, JK},
    journal={ECMWF Re-Analysis Profect Report Series},
    volume={72},
    year={1997}
}

@article{era40,
    author = {Uppala, Sai Sanmith and Kallberg, Per and Simmons, A and Andrae, Ulf and Bechtold, V and Fiorino, Mike and Gibson, J and Haseler, Jarmila and Hernandez-Carrascal, Angeles and Kelly, G and Li, X and Onogi, K and Saarinen, Sami and Sokka, N and Allan, Richard and Andersson, E and Arpe, Klaus and Balmaseda, Magdalena and Beljaars, Anton and Woollen, J},
    year = {2005},
    month = {01},
    pages = {2961-3012},
    title = {The ERA-40 reanalysis},
    volume = {131},
    journal = {Quarterly Journal of the Royal Meteorological Society}
}

@article{era-interim,
    title={The ERA-Interim reanalysis: Configuration and performance of the data assimilation system},
    author={Dee, Dick P and Uppala, S Mꎬ and Simmons, Adrian J and Berrisford, Paul and Poli, Paul and Kobayashi, Shinya and Andrae, U and Balmaseda, MA and Balsamo, G and Bauer, d P and others},
    journal={Quarterly Journal of the Royal Meteorological Society},
    volume={137},
    number={656},
    pages={553--597},
    year={2011},
    publisher={Wiley Online Library}
}

@article{roddier1979,
  title={The effects of atmospheric turbulence on the formation of visible and infrared images},
  author={Roddier, F},
  journal={Journal of Optics},
  volume={10},
  number={6},
  pages={299--303},
  year={1979}
}

@article{mahrt2013,
  title={Non-stationary generation of weak turbulence for very stable and weak-wind conditions},
  author={Mahrt, Larry and Thomas, Christoph and Richardson, Scott and Seaman, Nelson and Stauffer, David and Zeeman, Matthias},
  journal={Boundary-layer meteorology},
  volume={147},
  number={2},
  pages={179--199},
  year={2013},
  publisher={Springer}
}

@article{hellemeier2019,
  title={Weather at selected astronomical sites--an overview of five atmospheric parameters},
  author={Hellemeier, Joschua A and Yang, Rui and Sarazin, Marc and Hickson, Paul},
  journal={Monthly Notices of the Royal Astronomical Society},
  volume={482},
  number={4},
  pages={4941--4950},
  year={2019},
  publisher={Oxford University Press}
}

@article{fried1966,
    author = {Fried, David},
    year = {1966},
    month = {10},
    pages = {1372-1379},
    title = {Optical Resolution Through a Randomly Inhomogeneous Medium for Very Long and Very Short Exposures},
    volume = {56},
    journal = {J. Opt. Soc. Am.}
}

@article{xu2022,
    author = {Xu, Manman and Shao, Shiyong and Weng, Ningquan and Liu, Qing},
    year = {2022},
    month = {06},
    pages = {3085},
    title = {Analysis of the Optical Turbulence Model Using Meteorological Data},
    volume = {14},
    journal = {Remote Sensing}
}

@article{tatarskii1961,
    author = {Tatarski, V. I. and Silverman, R. A. and Chako, Nicholas},
    title = "{Wave Propagation in a Turbulent Medium}",
    journal = {Physics Today},
    volume = {14},
    number = {12},
    pages = {46-51},
    year = {1961},
    month = {12},
    issn = {0031-9228}
}

@article{Coulman1988,
    author = {C. E. Coulman and J. Vernin and Y. Coqueugniot and J. L. Caccia},
    journal = {Applied Optics},
    number = {1},
    pages = {155--160},
    publisher = {Optica Publishing Group},
    title = {Outer scale of turbulence appropriate to modeling refractive-indexstructure profiles},
    volume = {27},
    year = {1988}
    }

@inproceedings{ruggiero2002forecasting,
    title={Forecasting optical turbulence from mesoscale numerical weather prediction models},
    author={Ruggiero, Frank H and DeBenedictis, Daniel A},
    booktitle={DoD High Performance Modernization Program Users Group Conference},
    pages={10--14},
    year={2002}
}

@ARTICLE{hutt1999,
    author = {{Hutt}, Daniel L.},
    title = "{Modeling and measurement of atmospheric optical turbulence over land}",
    journal = {Optical Engineering},
    year = 1999,
    month = aug,
    volume = {38},
    pages = {1288-1295},
    doi = {10.1117/1.602188},
    adsurl = {https://ui.adsabs.harvard.edu/abs/1999OptEn..38.1288H}
}

@article{vernin1986,
    author = {Vernin, J.},
    year = {1986},
    month = {08},
    pages = {},
    title = {Astronomical Site Selection : A New Meteorological Approach},
    volume = {628},
    journal = {Proceedings of SPIE - The International Society for Optical Engineering},
    doi = {10.1117/12.963521}
}

@INPROCEEDINGS{sarazin2002,
    author = {{Sarazin}, Marc and {Tokovinin}, Andrei},
    title = "{The Statistics of Isoplanatic Angle and Adaptive Optics Time Constant derived from DIMM Data}",
    booktitle = {European Southern Observatory Conference and Workshop Proceedings},
    year = 2002,
    editor = {{Vernet}, E. and {Ragazzoni}, R. and {Esposito}, S. and {Hubin}, N.},
    series = {European Southern Observatory Conference and Workshop Proceedings},
    volume = {58},
    month = jan,
    pages = {321},
    adsurl = {https://ui.adsabs.harvard.edu/abs/2002ESOC...58..321S}
}

@inproceedings{gawad2006,
    author = {Gawad, Amal and Zoklot, Abdel-Salam and Abdel Gawad, Ahmed},
    year = {2006},
    month = {07},
    pages = {},
    title = {Wind and Environmental Effect on the Overhead High Voltage Transmission Lines},
    booktitle = {WIT Transactions on Ecology and the Environment (2006) },
    volume = {93},
    journal = {WIT Transactions on Ecology and the Environment},
    doi = {10.2495/SC060411}
}

@Article{bowen1977,
    AUTHOR = {Bowen, A. J. and Lindley, D.},
    TITLE = {A wind-tunnel investigation of the wind speed and turbulence characteristics close to the ground over various escarpment shapes},
    JOURNAL = {Boundary-Layer Meteorology},
    VOLUME = {12},
    YEAR = {1977},
    PAGES = {259-271},
    DOI = {https://doi.org/10.1007/BF00121466}
}

@article{li2022,
    doi = {10.1088/1674-4527/ac4e00},
    url = {https://dx.doi.org/10.1088/1674-4527/ac4e00},
    year = {2022},
    month = {mar},
    publisher = {National Astromonical Observatories, CAS and IOP Publishing},
    volume = {22},
    number = {4},
    pages = {045002},
    author = {Li, Tao-Ran and Jiang, Xiao-Jun},
    title = {Wind Environment Analysis of Ground-based Optical Observatory},
    journal = {Research in Astronomy and Astrophysics},
}

@article{fukao2003,
  title={Equatorial Atmosphere Radar (EAR): System description and first results},
  author={Fukao, Shoichiro and Hashiguchi, Hiroyuki and Yamamoto, Mamoru and Tsuda, Toshitaka and Nakamura, Takuji and Yamamoto, Masayuki K and Sato, Toru and Hagio, Masahiro and Yabugaki, Yoshiyuki},
  journal={Radio Science},
  volume={38},
  number={3},
  year={2003},
  publisher={Wiley Online Library}
}

@article{szkolka2025,
  title={Tropospheric winds over West Sumatra—a comparison between ERA-5 reanalysis and equatorial atmosphere radar},
  author={Szkolka, Wojciech and Baranowski, Dariusz B and Flatau, Piotr J and Marzuki and Shimomai, Toyoshi and Hashiguchi, Hiroyuki},
  journal={Climate Dynamics},
  volume={63},
  number={1},
  pages={88},
  year={2025},
  publisher={Springer}
}

@article{tonry2012,
  title={The Pan-STARRS1 photometric system},
  author={Tonry, JL and Stubbs, Christopher W and Lykke, Keith R and Doherty, Peter and Shivvers, IS and Burgett, WS and Chambers, KC and Hodapp, KW and Kaiser, N and Kudritzki, R-P and others},
  journal={The Astrophysical Journal},
  volume={750},
  number={2},
  pages={99},
  year={2012},
  publisher={IOP Publishing}
}

@book{wallace2006,
    author = {Wallace, J.M. and Hobbs, P.V.},
    year = {2006},
    month = {02},
    pages = {1-488},
    title = {Atmospheric Science: An Introductory Survey: Second Edition},
    address = {Singapore},
    publisher = {Elsevier},
}

@article{fairall1984,
      author = "C. W.  Fairall",
      title = "Wind Shear Enhancement of Entrainment and Refractive Index Structure Parameter at the Top of a Turbulent Mixed Layer",
      journal = "Journal of Atmospheric Sciences",
      year = "1984",
      publisher = "American Meteorological Society",
      address = "Boston MA, USA",
      volume = "41",
      number = "24",
      doi = "10.1175/1520-0469(1984)041<3472:WSEOEA>2.0.CO;2",
      pages= "3472 - 3484",
      url = "https://journals.ametsoc.org/view/journals/atsc/41/24/1520-0469_1984_041_3472_wseoea_2_0_co_2.xml"
}
\bibliographystyle{aasjournal}

\end{document}